\documentclass[aps,prl,superscriptaddress,showpacs, preprint
]{revtex4-1}
\usepackage{graphicx}
\usepackage{epstopdf}
\usepackage{amssymb,amsmath,amsfonts}

\newcommand{\cso} {{\mbox{Cu${}_2$OSeO${}_3$}}}
\begin{document}
\title{Critical scaling in a cubic helimagnet $\cso$}
\author{I. \v Zivkovi\'c}
\affiliation{Institute of Physics, Bijeni\v cka 46, HR-10000, Zagreb, Croatia}
\author{J. S. White}
\affiliation{Institute of Condensed Matter Physics, EPFL, CH-1015 Lausanne, Switzerland}
\affiliation{Laboratory for Neutron Scattering, Paul Scherrer Institut, CH-5232 Villigen,
Switzerland}
\author{H. M. R\o nnow}
\affiliation{Institute of Condensed Matter Physics, EPFL, CH-1015 Lausanne, Switzerland}
\author{K. Pr\v sa}
\affiliation{Institute of Condensed Matter Physics, EPFL, CH-1015 Lausanne, Switzerland}
\author{H. Berger}
\affiliation{Institute of Condensed Matter Physics, EPFL, CH-1015 Lausanne, Switzerland}

\date{\today}

\begin{abstract}
We present a detailed AC susceptibility investigation of the fluctuation regime in the insulating cubic helimagnet Cu$_{2}$OSeO$_{3}$. For magnetic fields $\mu_0 H \geq 200$ mT, and over a wide temperature ($T$) range, the system behaves according to the scaling relations characteristic of the classical 3D Heisenberg model. For lower magnetic fields the scaling is preserved only at higher $T$, and becomes renormalized in a narrow $T$ range above the transition temperature. Contrary to the well-studied case of MnSi, where the renormalization has been interpreted within the Brazovskii theory, our analysis of the renormalization at $H = 0$ shows the fluctuation regime in Cu$_{2}$OSeO$_{3}$ to lie closer to that expected within the Wilson-Fischer scenario.
\end{abstract}


%
%
%

\maketitle

%
%
%
%

Universality is a concept that lies at the heart of modern physics since it describes the general scaling behaviour of widespread physical phenomena within the vicinity of a critical point. In Condensed Matter physics, universal scaling laws are readily applied to interpret measurements of thermodynamic observables in order to discern the symmetry of the physical properties close to phase transitions. A classic example where these concepts have been extensively tested, both theoretically and experimentally, is the second-order paramagnetic (PM) to ferromagnetic (FM) transition~\cite{Wilson1983}. In general, as the system approaches the critical point, both the size and the number of fluctuations of the relevant order parameter increases. It is also known that if the interactions between the fluctuations are strong enough, that this may even alter the order of the phase transition. In a recent comprehensive study it was proposed~\cite{Janoschek2013} that a specific type of renormalization put forward by Brazovskii~\cite{Brazovskii1975} can be applied to describe the weakly first-order nature of the PM to helimagnetic (HM) transition at zero magnetic field in metallic MnSi~\cite{Pappas2009, Adams2012}. In this scenario, the renormalization arises due to the crucial role played by the Dzyaloshinskii-Moriya (DM) interaction which alters the nature of fluctuations close to $T_{HM}$, and so causes the system to avoid the second-order transition expected within mean field theory. Other studies of the unusual critical behaviour in MnSi include recent polarized neutron scattering experiments~\cite{Pappas2009}, from which it was proposed that a 'skyrmion-liquid' phase exists for a narrow temperature range ($T \sim 1$ K) above $T_{HM}$. A similar claim was deduced from a Monte Carlo study where an analogy with blue phases in liquid crystals has been established~\cite{Hamann2011}.

Recently, $\cso$ was identified as a new compound to display a direct PM to HM transition in zero magnetic field. $\cso$ crystallizes in the same space group as MnSi (P213), and has two crystallographically inequivalent Cu sites with a dominant antiferromagnetic interaction between the nearest neighbours~\cite{Yang2012}. The ratio of Cu ions within the two inequivalent sites is 3:1, giving rise to the formation of the local ferrimagnetic (FiM) 3-up-1-down state~\cite{Adams2012} which is then modulated by the DM interaction. Similar as for MnSi, by applying a weak magnetic field close to the ordering temperature, the magnetic moments in $\cso$ form a skyrmion lattice (SkL) which is a hexagonal arrangement of individual skyrmions -- whirls of spins –- that each have a non-zero topological charge~\cite{Adams2012,Seki2012}. In contrast to MnSi however, $\cso$ is a magneto-electric insulator. The magneto-electric coupling is caused by the $d-p$ hybridization mechanism, which explains the remarkable and hitherto unique polarization effects observed when the SkL is formed~\cite{Seki2012a}. It has been further suggested that an electric dipole can be assigned to an individual skyrmion~\cite{Seki2012a} and when applying an electric field the whole SkL has been observed to rotate~\cite{White2012}.

%
\begin{figure}
\includegraphics[width=0.45\textwidth]{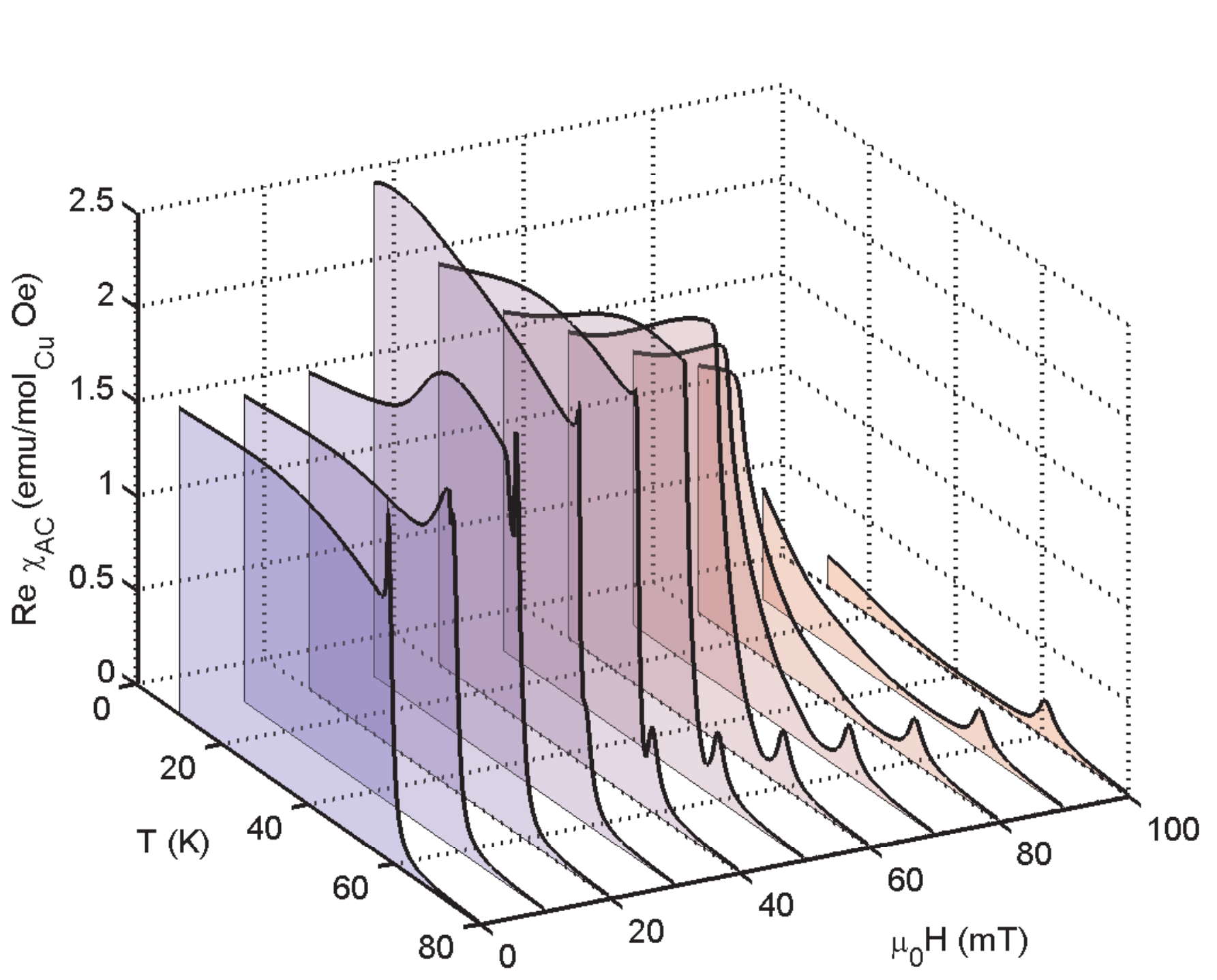}
\caption{(Color online) A set of temperature dependent ac susceptibility measurements of $\cso$ for magnetic fields up to 100 mT.}
\label{fig-overview}
\end{figure}
%

In this Letter we present the first study of the fluctuation region in $\cso$. A key result of our work concerns the experimental observation of the universal scaling at high magnetic fields (away from the zero field PM-HM transition), expected for any Heisenberg system. Our detailed AC susceptibility measurements show the existence of the scaling regime in a wide temperature and magnetic field range and provide an exceptional agreement with theory. On approaching the PM-HM transition at low magnetic fields we obtain another crucial result that the scaling is invalidated due to the influence of both the fluctuations and the DM interaction. We discuss the applicability of the Brazovskii renormalization for the case of $\cso$, and find that due to the increased strength of the interaction between the fluctuations, relative to those in MnSi, an extended description within a Wilson-Fischer scenario is necessary to describe the data.

%
%
%
%

The measurements have been performed on a home-made AC susceptibility setup with an excitation value of 0.1 mT and frequency of 1111 Hz. A DC magnetic field has been applied using a commercial 9 T superconducting magnet. The sample was a single crystal used in our previous work~\cite{Zivkovic2012} with dimensions $4 \times 1 \times 1$ mm$^3$, the longest dimension being along the [111] direction.

%
%
%
%

A set of temperature dependent susceptibility measurements with applied magnetic fields up to 100 mT is shown in Fig.~\ref{fig-overview}. In the range $0 \leq \mu_0 H \lesssim 30$ mT a first-order transition occurs at $\sim 58$ K. At $\mu_0 H = 20$ mT there is a narrow dip just below the transition, a hallmark of the SkL formation. Above 30 mT the transition becomes second-order~\cite{Adams2012} and the critical field quickly saturates towards lower temperatures. Above 90 mT one can notice a decaying contribution from the fluctuations of the conical order parameter within the FiM matrix.

%
\begin{figure}
\includegraphics[width=0.45\textwidth]{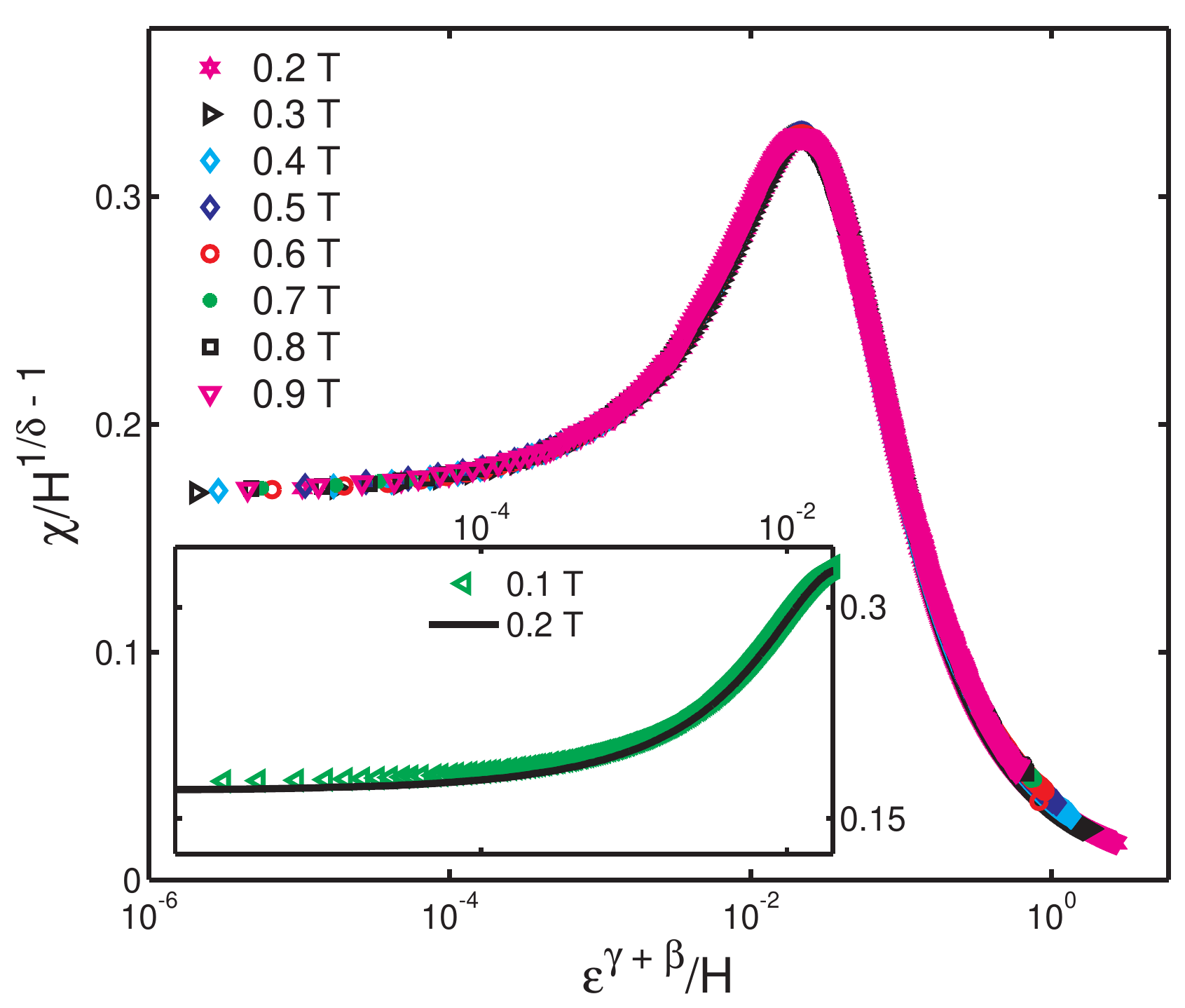}
\caption{(Color online) The analysis of the susceptibility measured in magnetic fields $\mu_0 H \geq 200$ mT according to the scaling hypothesis. The parameters used are $T_C = 58.3$ K, $\delta = 4.9$, $\beta = 0.37$ and $\gamma = 1.44$. Inset: Comparison of 100 mT and 200 mT scans very close to $T_C$.}
\label{fig-over2000}
\end{figure}
%

\begin{figure}
\includegraphics[width=0.45\textwidth]{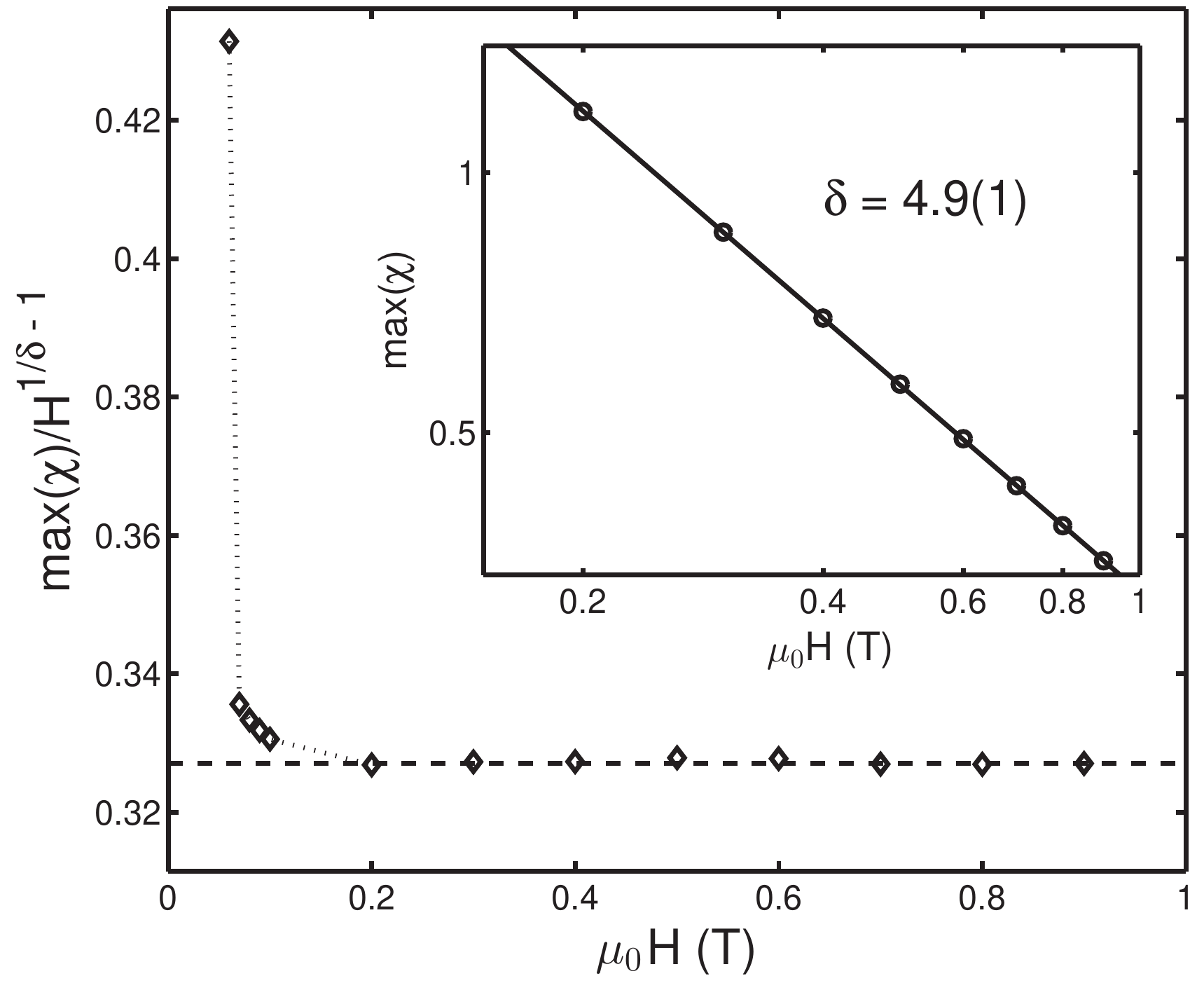}
\caption{Field dependence of the renormalized maximum of the susceptibility. Below 200 mT an increase is observed relative to the value above 200 mT indicated by the dashed line. The dotted line below 200 mT is a guide-to-the-eye. Inset: Log-log plot for fields above 200 mT.}
\label{fig-maximum}
\end{figure}

Around 60 K a broad feature can be noticed for magnetic fields above 30 mT. As the field is increased its magnitude decreases and it shifts slightly to higher temperatures. A similar feature was found in MnSi and other B20 compounds and interpreted as a smeared crossover between the high-$T$ non-polarized state and a low-$T$ field-polarized state~\cite{Thessieu1997}. In FeGe Wilhelm et \textit{al}.~\cite{Wilhelm2011} characterized the feature by its inflection points and argued that in the lower field region it transforms into skyrmion-like precursor states. Here, we point out that such a feature is characteristic of the classical ferromagnetic transition evolving with applied magnetic field. In this case a general scaling analysis~\cite{Stanley1975} of the type
%
\begin{equation}
\chi (T,H) \sim H^{1/\delta - 1} \mathcal{F} \left( \frac{H}{\varepsilon^{\gamma + \beta}} \right)
\label{eq-scaling}
\end{equation}
%
can be applied to confirm such a behaviour and to extract critical exponents. Here, $\varepsilon = T/T_C - 1$ and $\mathcal{F} \left( H/(\varepsilon^{\gamma + \beta}) \right)$ is a scaling function for $T > T_C$. In order to analyse the critical behaviour, we have measured $\chi (T)$ up to 100 K in the field range $0 \leq \mu_0 H < 1$ T. In Fig.~\ref{fig-over2000} we present the results of such scaling, which seems to work exceptionally well for $\mu_0 H \geq 200$ mT. At lower fields a gradual influence of the conical order parameter can be noticed. This can be visualized by plotting the field dependence of the maximum of the normalized susceptibility $\chi/H^{1/\delta -1}$ (see Fig.~\ref{fig-maximum}) for which the scaling hypothesis predicts a simple relation $\max(\chi) \propto H^{1/\delta - 1}$. Below 200 mT the maximum starts to increase relative to the level obtained for larger fields (the dashed line). The value of $\delta = 4.9(1)$ has been extracted from the log-log plot displayed in the inset of Fig.~\ref{fig-maximum}.
%
%
%
%
This leaves only two unknown parameters since the scaling relation $\gamma = \beta (\delta - 1)$ applies. An excellent data collapse is obtained using $T_C = 58.3(1)$ K and $\beta = 0.37(1)$, which yields $\gamma = 1.44(4)$, in excellent agreement with the theoretical exponents for the 3D-Heisenberg universality class ($\beta = 0.365$, $\gamma = 1.39$, $\delta = 4.8$)~\cite{Chen1993}. The critical temperature $T_C$ obtained from the scaling is very close to the HM transition temperature $T_{HM} = 58.1$ K for $\mu_0 H = 0$ T (see Fig.~\ref{fig-brazovskii} below).

It is important to point out that the scaling is valid over several decades of the reduced temperature ($\varepsilon \lesssim 10^{0}$) and in a wide range of magnetic fields. Remarkably, the microscopic details that give rise to the ferrimagnetic component in $\cso$ (the 3-up-1-down configuration) do not seem to play an important role in the investigated ($H$,$T$) range.

%
\begin{figure*}
\includegraphics[width=0.9\textwidth]{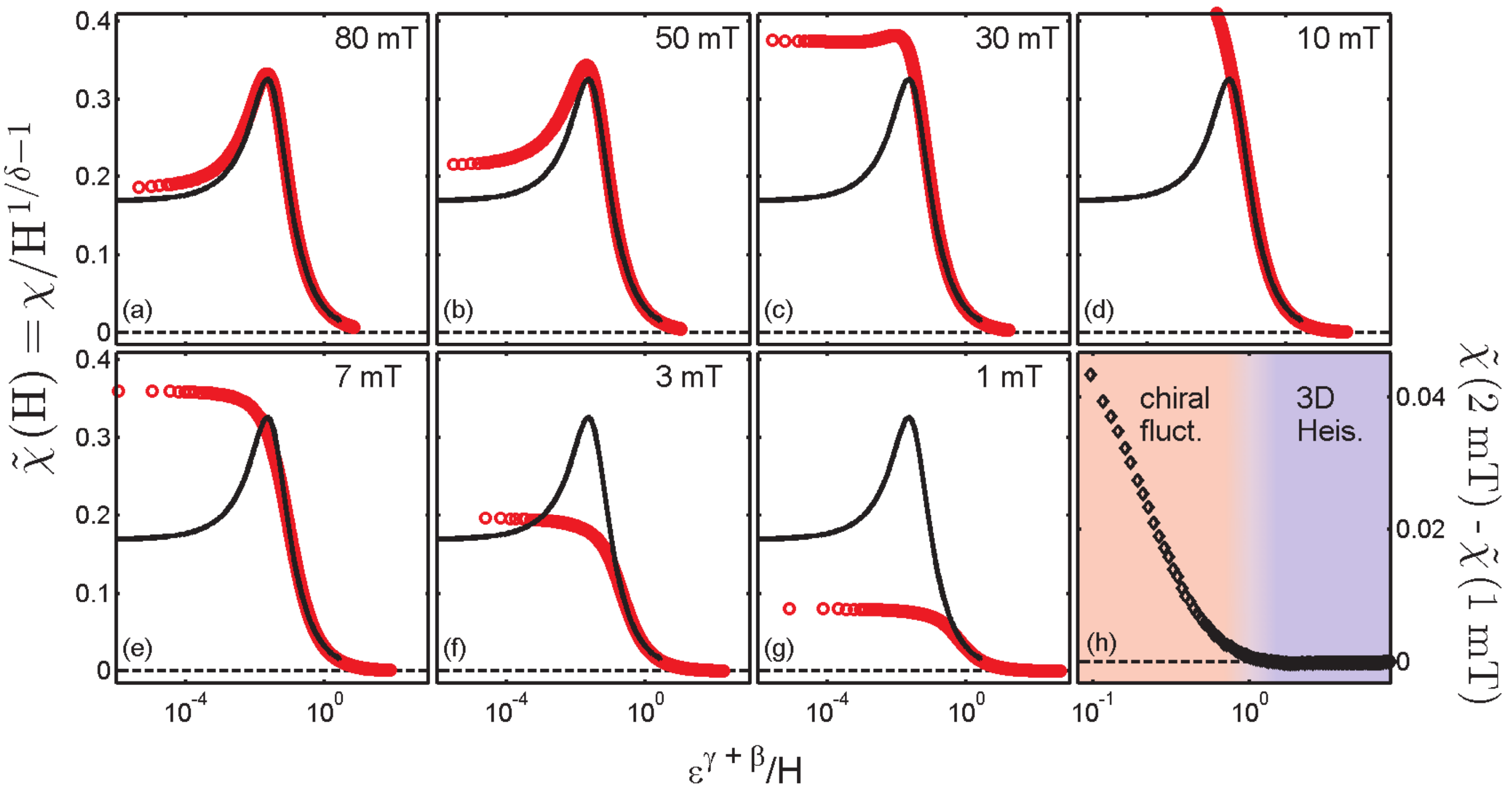}
\caption{(Color online) (a)--(g) Evolution of the divergence from the critical behaviour for fields below 100 mT. The solid line shows the scaling behaviour for $\mu_0 H = 200$ mT. (h) The difference plot between the FM critical behaviour and the measurement at $\mu_0 H = 1$ mT.}
\label{fig-evolution}
\end{figure*}
%

We continue now with the investigation of the susceptibility below 200 mT. As the field is lowered the fluctuations of the conical order parameter start to build up at higher temperatures, following the line of the second-order transition to the ordered state. Around $\mu_0 H \approx 100$ mT the difference between the measured curve and the ideal FM critical behaviour (represented by the $\mu_0 H = 200$ mT curve) is small and noticeable only when zoomed-in for temperatures close to $T_C$ (see the inset of Fig.~\ref{fig-over2000}). At the same time, on the high-$T$ side the scaling is still obeyed (not shown). When the field is further decreased the difference becomes more pronounced in the temperature range above $T_C$ but below the maximum of the FM critical curve, as shown in Fig.~\ref{fig-evolution}(a)-(c). In the field range $\mu_0 H \leq 10$ mT, the curves also begin to deviate from the ideal FM critical behaviour on the high-$T$ side of the maximum. However, the divergence becomes qualitatively different since the measured susceptibility is now below the ideal FM critical behaviour, indicating a situation where the criticality is avoided~\cite{Zivkovic2012}. Although the susceptibility above $T_C$ does not change much as $H \rightarrow 0$ (Fig.~\ref{fig-overview}), the ideal FM critical behaviour rapidly diverges, resulting in the relative decrease of the normalized susceptibility value just above $T_C$ (Fig.~\ref{fig-evolution}(e)-(g)).

For each value of $H$ we can estimate the temperature where the divergence becomes pronounced. This is done by subtracting the normalized susceptibility $\tilde{\chi} (H) = \chi/H^{1/\delta - 1}$ from the ideal FM critical behaviour obtained from the normalized susceptibility at some higher field value. The case for $\mu_0 H = 1$ mT is shown in Fig.~\ref{fig-evolution}(h). By this approach we could estimate the cross-over temperature for fields up to $\mu_0 H = 40$ mT, above which the divergence starts to approach the maximum of the scaling curve and the difference becomes less pronounced, rendering the determination ambiguous. We have found that for all the curves the cross-over occurs around 60 K.

Qualitatively the same avoidance of the criticality has been observed in MnSi and other B20 compounds~\cite{Janoschek2013,Grigoriev2011,Bauer2010}. Several interpretations have been put forward, including skyrmion liquid phase~\cite{Pappas2009}, magnetic blue phase~\cite{Hamann2011}, chiral fluctuating states~\cite{Grigoriev2011} and the Brazovskii renormalization scenario~\cite{Janoschek2013}. Janoschek and co-workers argued that the Brazovskii theory~\cite{Brazovskii1975} works not only for MnSi but also for other HM systems governed by the DM interaction~\cite{Janoschek2013}. In what follows we discuss its applicability on $\cso$.

On a mean-field level, the HM transition is expected to be second order. However, strong interactions between fluctuations of the order parameter drive the transition first order, with the strength of the interaction defining a characteristic length scale $\xi_G$ (Ginzburg scale). This leads to the suppression of the transition temperature and to the renormalization of the temperature dependence of the correlation length $\xi (T)$. It has been shown~\cite{Janoschek2013} that within the Brazovskii scenario the longitudinal susceptibility for $T > T_{HM}$ can be expressed as
%
%
\begin{equation}
\chi (T) = \frac{\chi_0}{1 + \eta ^2 \mathcal{Z}^2 \\(T)}
\label{eq-renormalized}
\end{equation}
%
%
with $\eta = \xi_{DM} / \xi_G$ and $\mathcal{Z} (T)$ given by
%
%
\begin{equation}
\mathcal{Z} (T) = \frac{\sqrt[3]{2} + (1 + \sqrt{1 - 2\tau ^3})^{1/3}}{\sqrt[3]{2}(1 + \sqrt{1 - 2\tau ^3})^{1/3}}
\label{eq-brazovskii}
\end{equation}
%
%
where $\tau = (T - T_{MF})/T_0$ is the reduced temperature relative to the mean-field value $T_{MF}$. For temperatures close enough to $T_{HM}$ $\xi (T) > \xi_{DM}$ and the DM interaction becomes important, causing the fluctuations to acquire a chirality.

It has been suggested that at high temperatures, where $\xi (T) \ll \xi_{DM}$, the system recovers the mean-field behaviour~\cite{Janoschek2013}. We have shown that $\cso$ follows the critical scaling relation even up to $\varepsilon \approx 1$. The crucial question is, in what temperature range does Eq.~(\ref{eq-renormalized}) apply? To address this issue we have performed a series of fits starting from 58.3 K (just above the maximum of the measured curve) up to a variable upper bound $T^{up}$. The inset of Fig.~\ref{fig-brazovskii} shows the goodness-of-fit given by $RMS^2 = \sum\limits_{i = 1}^{N} (\chi_{calc} (T_i) - \chi_{meas} (T_i))^2 / N)$. Here, a division with the number of points $N$ normalizes the variability in the fitting range. A pronounced minimum in $RMS^2$ is seen, and the best fit is obtained for $T^{up} = 60.14$ K. This value is in good agreement with the cross-over temperature obtained from the splitting of the low-$H$ curves and the FM scaling behaviour (Fig.~\ref{fig-evolution}(h)).

%
\begin{figure}
\includegraphics[width=0.45\textwidth]{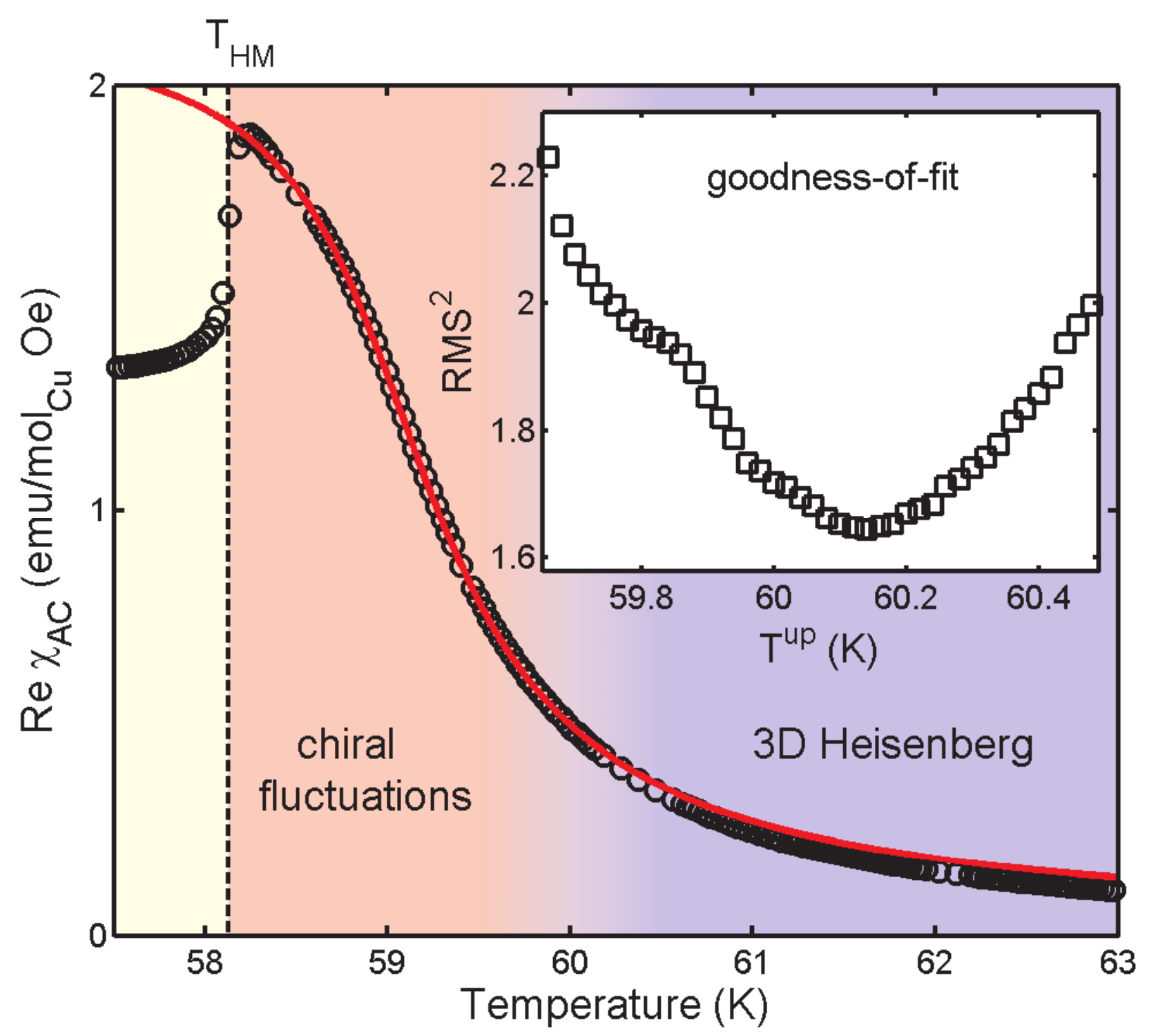}
\caption{(Color online) The temperature dependence of the susceptibility for $\mu_0 H = 0$ T. The solid line is a fit to Eq.~(\ref{eq-brazovskii}). Inset: the goodness-of-fit vs.~the upper bound of the fit.}
\label{fig-brazovskii}
\end{figure}
%

The main panel of Fig.~\ref{fig-brazovskii} shows the comparison between the best fit and the measured data. A very good agreement exists close to $T_{HM}$ but at higher temperatures two curves start to diverge. The parameters obtained from the fit are $\chi_0 = 6.82(3)$, $\eta = 1.185(11)$, $T_{MF} = 59.39(1)$ K and $T_0 = 0.864(8)$ K. Importantly, in the temperature range considered in the inset of Fig.~\ref{fig-brazovskii} the exact choice of the upper bound of the fit does not influence their values considerably (only a couple of percent).

%
%
%
%

The Brazovskii scenario implies $\xi_{DM} \ll \xi_G $ which has been demonstrated to work for MnSi ($\eta = \xi_{DM} / \xi_G \approx 0.5$)~\cite{Janoschek2013}. For $\cso$ we have obtained $\xi_{DM} \gtrsim \xi_G$ which indicates that in the context of the renormalization theory a Wilson-Fisher scenario must be taken into account~\cite{Wilson1983}, where strong interactions between the fluctuations arise before the DM interaction introduces the chirality in the system. However, it might be expected that other interpretations for MnSi~\cite{Pappas2009,Hamann2011,Grigoriev2011} could give a slightly different renormalization of the susceptibility which could be tested against detailed temperature and magnetic field dependences in $\cso$. Such a systematic approach could complement investigations done using neutron scattering experiments.

Our study of the scaling behaviour further reveals a remarkable change in the nature of the divergence close to 10 mT, see Fig.~\ref{fig-evolution}(a-g). For $\mu_0 H < 10$ mT the measured susceptibility diverges from the expected FM critical behaviour on the lower side while for $\mu_0 H > 10$ mT it goes above it. Around the same value of 10 mT the SkL phase begins to appear in the ordered region of $\cso$, suggesting that the triple-$Q$ structure~\cite{Muhlbauer2009} might be related to the change in the type of the divergence the system is following. The investigation of the fluctuation region in thin films, where a much broader SkL region is observed, would help clarify this issue.

%
%
%
%

In conclusion, detailed AC susceptibility measurements have revealed a broad feature to exist just outside the long-range ordered magnetic phase diagram of $\cso$. Our analysis shows this feature to follow a scaling hypothesis with critical exponents in remarkable agreement with a 3D Heisenberg model, and over a wide range of temperature and magnetic field. At very low magnetic fields, the deviation from the scaling observed close to the transition indicates that compared with MnSi, $\cso$ exhibits a stronger interaction between fluctuations of the order parameter. Such behaviour requires a theoretical approach within the Wilson-Fischer scenario which, to the best of our knowledge, is still not fully developed. This marks out $\cso$ as an ideal model system for testing new scaling theories.

%
%
%
%

The support from the Croatian Science Foundation project $02.05/33$ and the Swiss NCCR and its programme MaNEP are acknowledged.

\bibliography{Cu2OSeO3}

\end{document}